\documentclass[journal=jacsat,manuscript=article]{achemso}
\usepackage{chemformula}
\usepackage[T1]{fontenc}
\usepackage[utf8]{inputenc}
\usepackage{xspace}
\usepackage{amsmath}
\usepackage{threeparttable}
\usepackage{adjustbox}
\usepackage{amssymb}
\usepackage{graphicx}
\usepackage{subfigure}
\usepackage{xcolor}

\usepackage{orcidlink}

\usepackage{titlesec}
\titleformat{\section}{\normalfont\Large\bfseries}{\thesection}{1em}{}
\titleformat{\subsection}{\normalfont\large\bfseries}{\thesubsection}{1em}{}

\DeclareUnicodeCharacter{2212}{-}

\author{Anuroopa Behatha, \orcidlink{0000-0001-8911-8862}}
\author{Shalini Tomar, \orcidlink{0000-0003-2656-2928}}
\author{Hojin Jeong, \orcidlink{0000-0003-0825-3427}}
\author{Joon Hwan Choi}
\affiliation{Functional Ceramics Department
Korea Institute of Materials Science (KIMS)
797 Changwon-daero, Seongsan-gu, Chanwon-si, Gyeongsangnam-do, 51508, South Korea}
\email{jchoi@kims.re.kr}
\author{Seung-Cheol Lee, \orcidlink{0000-0002-9741-6955}}
\email{leesc@kist.re.kr}
\affiliation[First University]{Indo-Korea Science and Technology Center (IKST), Bangalore 560064, India}
\alsoaffiliation[Second University]
{Electronic Materials Research Center, Korea Institute of Science \& Technology, Seoul 130-650, Korea}
\author{Satadeep Bhattacharjee, \orcidlink{0000-0002-6717-2881}}
\email{s.bhattacharjee@ikst.res.in}
\affiliation[Unknown University]
{Indo-Korea Science and Technology Center (IKST), Bangalore 560064, India}

\title{Mechanistic Insights into Complete Methane Oxidation on Single-Atom Pd Supported by SSZ-13 Zeolite: A First-Principles Study}


\begin{document}

\begin{abstract} 
Complete catalytic oxidation of methane is an effective strategy for greenhouse gas mitigation and clean energy conversion; yet, ensuring both high catalytic activity and stability with palladium-based catalysts remains a challenge. In the present work, we employed a theoretical investigation of methane oxidation over single-atom Pd supported on SSZ-13 zeolite using density functional theory calculations, combined with climbing-image nudged elastic band calculations to determine activation barriers. A systematic assessment of various Al distributions and Pd placements was carried out to identify the most stable configurations for Pd incorporation within the zeolite framework.Further, two mechanistic routes for methane activation were evaluated: (i) direct dehydrogenation under dry conditions, and (ii) O$_2$-assisted oxidative dehydrogenation. Our results demonstrate that the direct (dry) pathway is energetically demanding and overall endothermic, whereas the O$_2$ assisted route facilitates the exothermic energy profile, particularly in the C-H bond cleavage. The formation of stable hydroxyl and CO/CO$_2$ intermediates were also studied. The results emphasize the role of oxygen-rich environments in enabling the complete methane oxidation with improved thermodynamic feasibility. Moreover, we propose an alternate low-energy pathway based on O-assisted and multi-site mechanisms that reduce the overall reaction enthalpy. These insights provide the design principles for highly active and moisture-resistant Pd-zeolite catalysts for sustainable methane utilization.
\end{abstract}

\section{Introduction}
\label{sec:intro}
The global warming crisis, fueled by the excessive greenhouse gas emission has become a focal point of international concern. Methane, an earth-abundant gas and a primary component of natural gas has been regarded as a high-efficient fuel \cite{ravi2019misconceptions}. However, utilization of methane as a fuel source imposes economic penalty \cite{zhang2024review} which highly recommends the conversion of methane into useful chemical products. The complete catalytic oxidation of methane to CO$_2$ substantially diminishes the environment impact by 96\%, as methane has a global warming potential of approximately 28-36 times greater than that of CO$_2$ over a 100-year timeline \cite{epa2021understanding}. This catalytic combustion of methane represents a flameless combustion process that lowers the ignition temperature of fuel, enhances the combustion efficiency of methane, and reduces the generation of atmospheric pollutants. Hence, it relies on many factors such as size, coordination geometry of the catalyst, nature of support, and many more \cite{Chen2023, Chacko2025, Escandon2005-zs}.
Palladium-based catalysts have been extensively studied for complete methane oxidation, typically dispersed on various high-surface-area metal oxide supporters. Notably, the  Pd/Al$_2$O$_3$ has received significant interest over the years \cite{GELIN20021,CHOUDHARY20021,monai2018catalytic}. The critical aspect of the methane oxidation catalysts is the selection of the support materials. In contrast to the high-surface-area metal oxides such as Al$_2$O$_3$, micro-porous zeolite offers numerous advantages including hydrophobic properties \cite{wang2020situ,zhang2019anti} precise control over the structure and composition and many more. In addition, Pd tends to reduce more easily over the zeolite compared to Al$_2$O$_3$, although it exhibits lower initial activity under both dry and wet conditions \cite{losch2019modular}. Despite some challenges related to steaming and de-alumination, recent works suggest that zeolite-based catalysts offer more affinity to water than others. Apart from that, zeolites are often stable at elevated temperatures.
With the aforementioned facts, zeolite-based catalysts are strongly believed to provide a potential platform in industrial applications for their exceptional catalytic activities and product selectivities \cite{shamzhy2019new, sun2021advances} compared to many well-known supporters such as  AlO$_2$, SiO$_2$ and ZrO$_2$ \cite{Ibrahim2020-ml, Chen2014, Liu}. For instance, Pd/Zeolite catalysts are by far the most commonly used catalysts for the complete oxidation of methane with excellent catalytic performance \cite{CHEN2022118534} 
In the current work, we perform a thorough theoretical investigation of methane oxidation over single-atom Pd supported on the SSZ-13 zeolite framework using density functional theory (DFT) calculations. Pd was chosen for its  proven ability to activate C–H bonds and its key role in methane combustion catalysis, while the SSZ-13 zeolite framework provides a highly stable and tunable support with superior hydrothermal resistance, thereby positioning the Pd@SSZ-13 system a strong contender for real-world applications. As a first step, we examined the incorporation of Pd at various framework positions associated with different Al distributions, in order to identify the most stable Pd–Al configurations and thereby determine the most probable active sites. Next, we investigated the mechanistic sequence of elementary reactions involved in methane oxidation under two different conditions: (i) the direct (dry) dehydrogenation pathway and (ii) the O$_2$-assisted oxidative dehydrogenation pathway. The role of oxygen in lowering activation barriers and enhancing the overall exothermicity of the reaction was systematically assessed. Activation energy barriers were estimated using the climbing image nudged elastic band (CI-NEB) method. Overall, our results reveal that the catalytic activity and thermodynamic feasibility of methane oxidation are strongly governed by the local environment of Pd within the zeolite framework and the availability of oxygen species, providing new insights for the design of efficient Pd/zeolite catalysts for methane combustion.

\section{Computational Details}
First-principles-based spin-polarized calculations were performed out using Vienna Ab-initio Simulation Package\textsc{VASP})\cite{Kresse1993,Kresse1996,Kresse1999}.  Exchange-correlation was treated within the generalized gradient approximation (GGA) using Perdew-Burke-Ernzerhof (PBE) functional \cite{Perdew1996_PBE1, Perdew1997_PBE2,Perdew2008}. Valence electrons considered in this calculation include Si: s2p2 (version 05Jan2001), O: s2p4 (version 08Apr2002), and Al: s2p1 (version 04Jan2001). A plane-wave energy cutoff of 500 eV was adopted for all the calculations and the Brillouin zone was sampled at the $\Gamma$ point. The total energy and interatomic forces were converged to 10$^{-6}$ and 0.001 eV/\AA. Adsorption energy has been computed using $$E_{ads} = E_{system+adsorbate} - E_{system} - E_{adsorbate}$$ Enthalpy of each reaction step has been calculated using $\Delta_H = E_p -Er $, where $E_p$ \& $E_r$  represents energy of products and reactants respectively. Activation energy barriers were estimated using the climbing image nudged elastic band (CI-NEB) method \cite{NEB1, NEB2}. To complement the CI-NEB calculations, we employed the Brønsted–Evans–Polanyi (BEP) relation and the Unity Bond Index–Quadratic Exponential Potential (UBI-QEP) method for estimating activation barriers \cite{BLIGAARD2004206, SHUSTOROVICH19981, C9CP04286E, doi:10.1021/acscatal.1c04347}.
In the BEP framework, activation energies (E$_a$) were expressed as a linear function of reaction energies $\Delta E = \gamma \Delta E + \xi E_a$ 
where family-specific slopes ($\gamma$) were adopted from literature, and the intercept ($\xi$) was calibrated to the NEB-computed CH$_4$ activation barrier. For the UBI-QEP (Shustorovich formalism) \cite{SHUSTOROVICH19981}, barriers were estimated as a function of gas-phase bond dissociation and surface adsorption energies. A simplified $\phi$-tuned form was employed: $E_a = \phi \Delta E + \xi_\phi E_a$
with $\phi = 0.85$ following prior benchmarking studies, and $\xi_\phi$ anchored to the CH$_4$ NEB barrier.
This combined strategy enabled barrier estimation for rate-determining steps along both oxygen-rich and oxygen-lean pathways while reducing computational cost relative to exhaustive NEB calculations. Bader analysis algorithm had been used to calculate the atomic spin densities\cite{sanville2007improved, HENKELMAN2006354}. Finally, the \textsc{vesta} code \cite{Momma2011} was utilized to visualize all the optimized structures.

\section{Results and Discussion}
\subsection{Active site and Structural details}
\begin{figure*}[t]
\centering
\includegraphics[width=14.0cm,keepaspectratio=true]{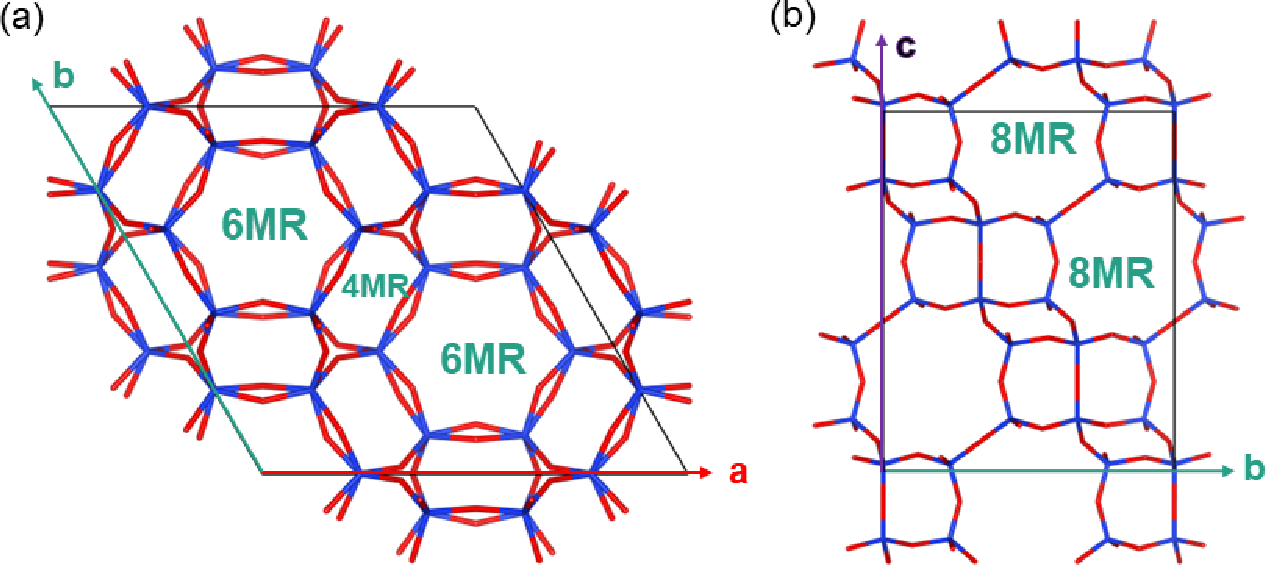}
\centering
\caption{(Color online) Zeolite (SSZ-13) framework highlighting the 6-,4-MR in [001] and (b) 8MR in [100] direction. Here, red and blue sticks represent O and Si atoms, respectively.}
\label{SSZ-13_str}
\end{figure*}

The crystal structure of SSZ-13 Zeolite consists of a three-dimensional framework of  Si, Al, and O atoms that belong to the chabazite (CHA) zeolite family \cite{zones1985zeolite} as shown in Fig. \ref{SSZ-13_str}.  The structure consists of  4-, 6- and 8-membered ring (MR) channels along 100, 001 directions. The detailed structure is given in supporting information (SI) Fig. S1. Two equivalent SSZ-13 unit cells can be constructed, including the hexagonal unit cell with 36 symmetry-equivalent tetrahedra (T) sites and 72 O atoms and the rhombohedral unit cell with 12T sites and 24 O atoms. However, in our present work, we have considered the hexagonal unit cell with lattice parameters, ( a =  13.675 A ;  c = 14.767 A). 

The investigation of potential active sites on Pd@SSZ-13 was carried out using a Si/Al ratio of 17 within zeolite framework. While recent experimental studies employ SSZ-13 with a Si/Al ratio of 15 \cite{Ryu2024}, achieving this value theoretically requires large supercell, which is computationally demanding. To maintain computational feasibility while remaining consistent with experimental conditions, we modelled the system with a Si/Al ratio of 17 in the unit cell. This was achieved by replacing two 12 T-site Si atoms with two Al atoms, with all arrangements being charge compensated by adding 2 background electrons \cite{Walker}. 

\begin{figure*}[t]
\centering
\includegraphics[width=14.0cm,keepaspectratio=true]{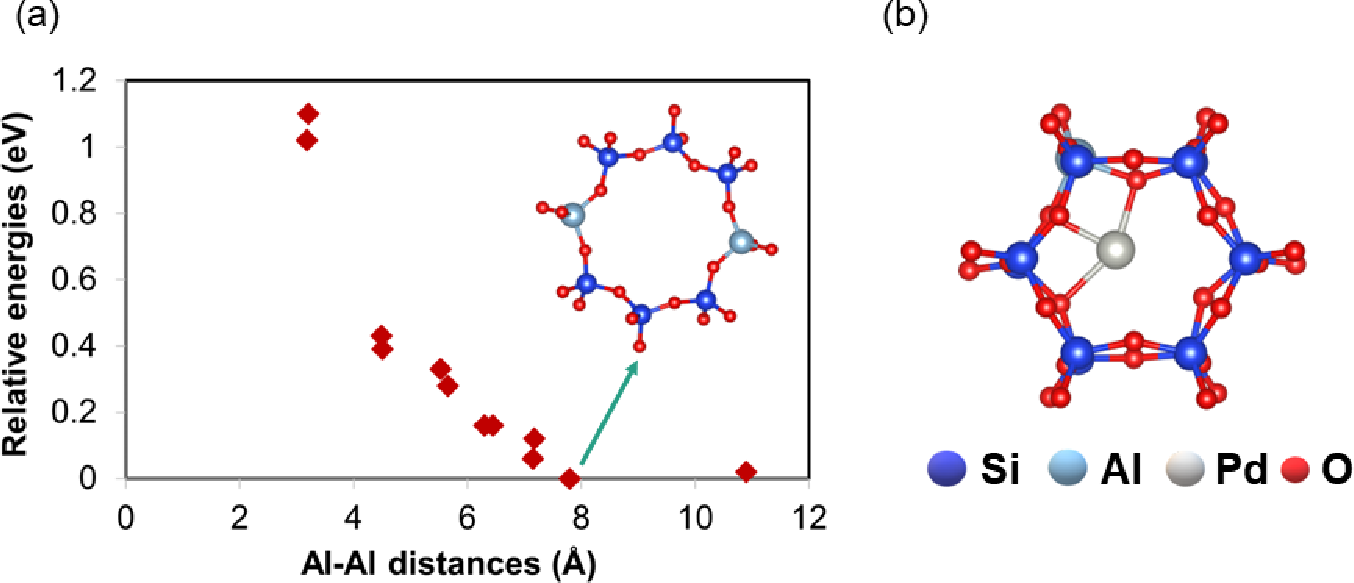}
\centering
\caption{(Color online) (a)Relative energies of various dual-alluminum configurations in SSZ-13 as a function of Al-Al seperation [here, energies are referenced to the lowest energy structure] (b) Top-view illustration of a Pd atom within the 6- MR of the Zeolite framework}
\label{Al_arrangements}
\end{figure*}

Further, we have proceeded with the geometry optimization of different Al arrangements followed by the total energy calculations. Fig. \ref{Al_arrangements}a provides the details of relative energies of different Al-substituted structures as a function of Al-Al distances. We observe that as the distance between aluminum atoms increases, the Zeolite framework becomes more stable. This means that the aluminum pairs that are three or four neighbors (T-sites) away from each other are the most suitable positions within SSZ-13. This finding is consistent with the previous experimental observation by Mlekodaj et al. \cite{Mlekodaj_exp} and theoretical observation by Eric A. Walker et. al \cite{Walker}.

Furthermore, the impact of palladium on the  SSZ-13 zeolite has been studied by placing Pd at the center of 4-, 6- ,and 8-MR respectively. This analysis provides insight into the preferred zeolite environment for palladium placement. We observe that, when a palladium atom is placed within a 6-MR (see Fig. \ref{Al_arrangements}b), the structure becomes more stabilized when compared to that of the 4-MR and 8-MR environments \cite{Aljama2022-zd, Kaushik}. This is possibly due to the ease with which palladium atoms coordinate with three framework oxygens, resulting in lower relative energy.
In this study, we classified the complete methane oxidation mechanism into two distinct categories based on reaction conditions: dry condition and wet/O$_2$-assisted condition. Each category encompasses three fundamental steps: methane dehydrogenation, O$_2$ dissociation, and the formation of CO$_2$ and H$_2$O.
\subsection{Proposed Reaction pathways}
Under dry condition, methane oxidation over Pd@SSZ-13 proceeds via sequential dehydrogenation, ultimately forming elemental carbon (C$^*$) and hydrogen (H$^*$). The adsorbed O$_2$ molecule dissociates into two atomic oxygen species (O$^*$) on the Pd surface. Subsequently, CO$_2$ is generated via the reactions $CO^* + O^* \rightarrow CO_2^*$, while water forms via stepwise hydrogen oxidation reactions: $O^* + H^* \rightarrow OH^*$ and $OH^* + H^* \rightarrow H_2O$. 
Further, two mechanistic routes were examined: (i) an oxygen-rich pathway (oxidant-excess/humid), proceeding via a deep C$^*$ route before complete oxidation to CO$_2$ and H$_2$O, and (ii) an oxygen-lean pathway (low O$_2$), proceeding via aldehyde-type intermediates (CH$_2$O$^*$, CHO$^*$) and avoiding direct C$^*$ formation. Both routes involve the same overall oxidation chemistry, but differ in key intermediates and activation barriers, as analyzed using BEP-anchored \cite{BLIGAARD2004206} and UBI-QEP methods \cite{SHUSTOROVICH19981}.
\phantomsection
\subsubsection{Dry condition}
In Dry condition, methane oxidation over Pd@SSZ-13 begins with stepwise C-H bond cleavage (direct dehydrogenation) from the CH$_4$ molecule, producing surface-bound C* species,  O$_2$ dissociation, which is followed by  CO$_2$, H$_2$O formation. CH$_4$ is weakly physisorbed with negligible charge transfer, whereas the other species like CH$_3$, CH$_2$, CH and C shows chemisorption with a charge transfer from Pd to C. The stronger interactions result in a higher activation barriers.  

\begin{figure*}[t]
\centering
\includegraphics[width=14.0cm,keepaspectratio=true]{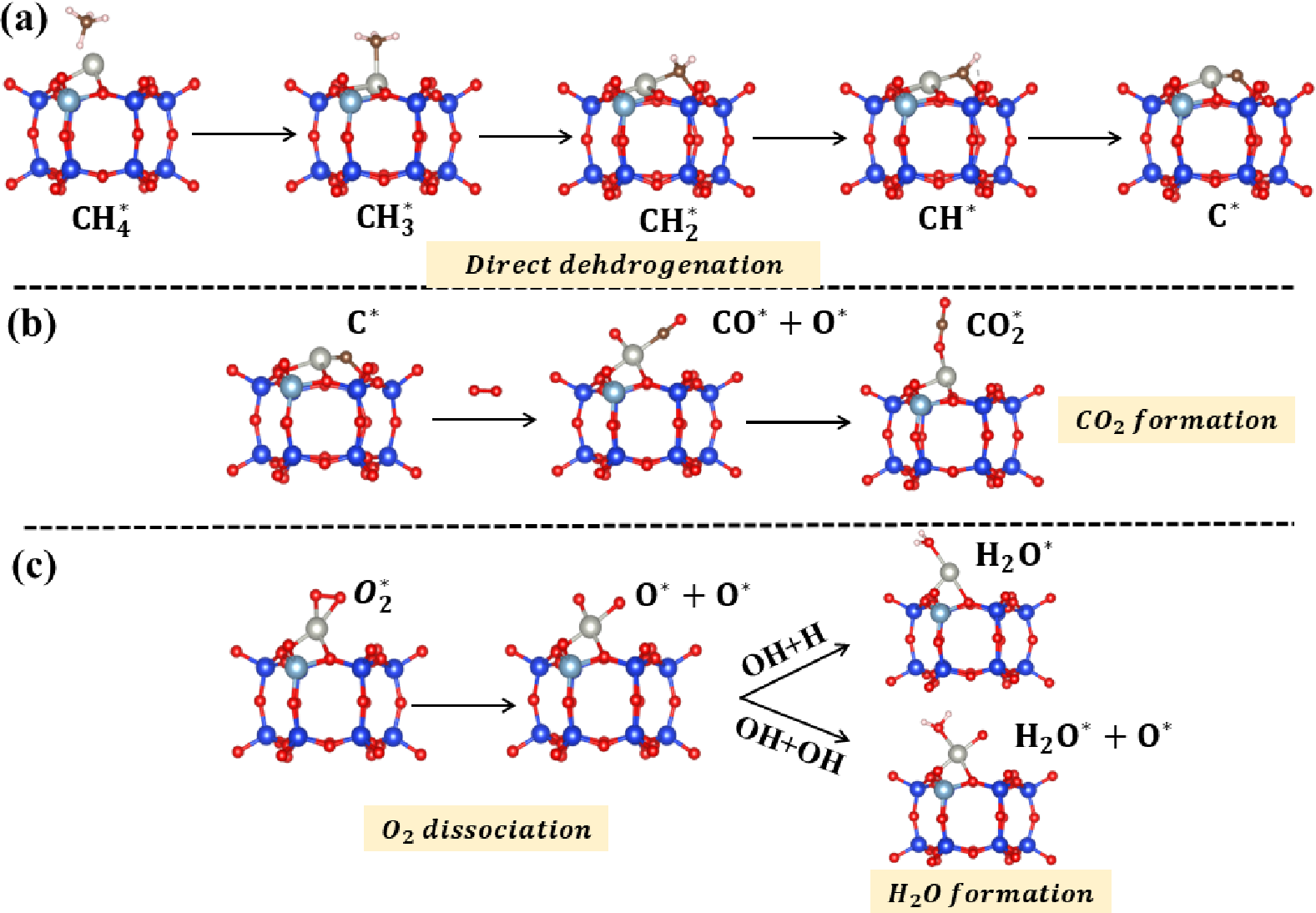}
\centering
\caption{(Color online) Side-view representations of the optimized adsorption geometries (a) CH$_4^*$, CH$_3^*$, CH$_2^*$, CH$^*$, C$^*$ species corresponding to the \textit{direct dehydrogenation} (b) C$^*$, CO$^*$+O$^*$, CO$_2^*$ intermediates involved in CO$_2$ formation (c) O$_2^*$, dissociated O$^*$+O$^*$, H$_2$O$^*$, H$_2$O$^*$+O$^*$ configurations associated with O$_2$ activation and H$_2$O formation)}
\label{Str_dry}
\end{figure*}

\begin{figure*}[t]
\centering
\includegraphics[width=14.0cm,keepaspectratio=true]{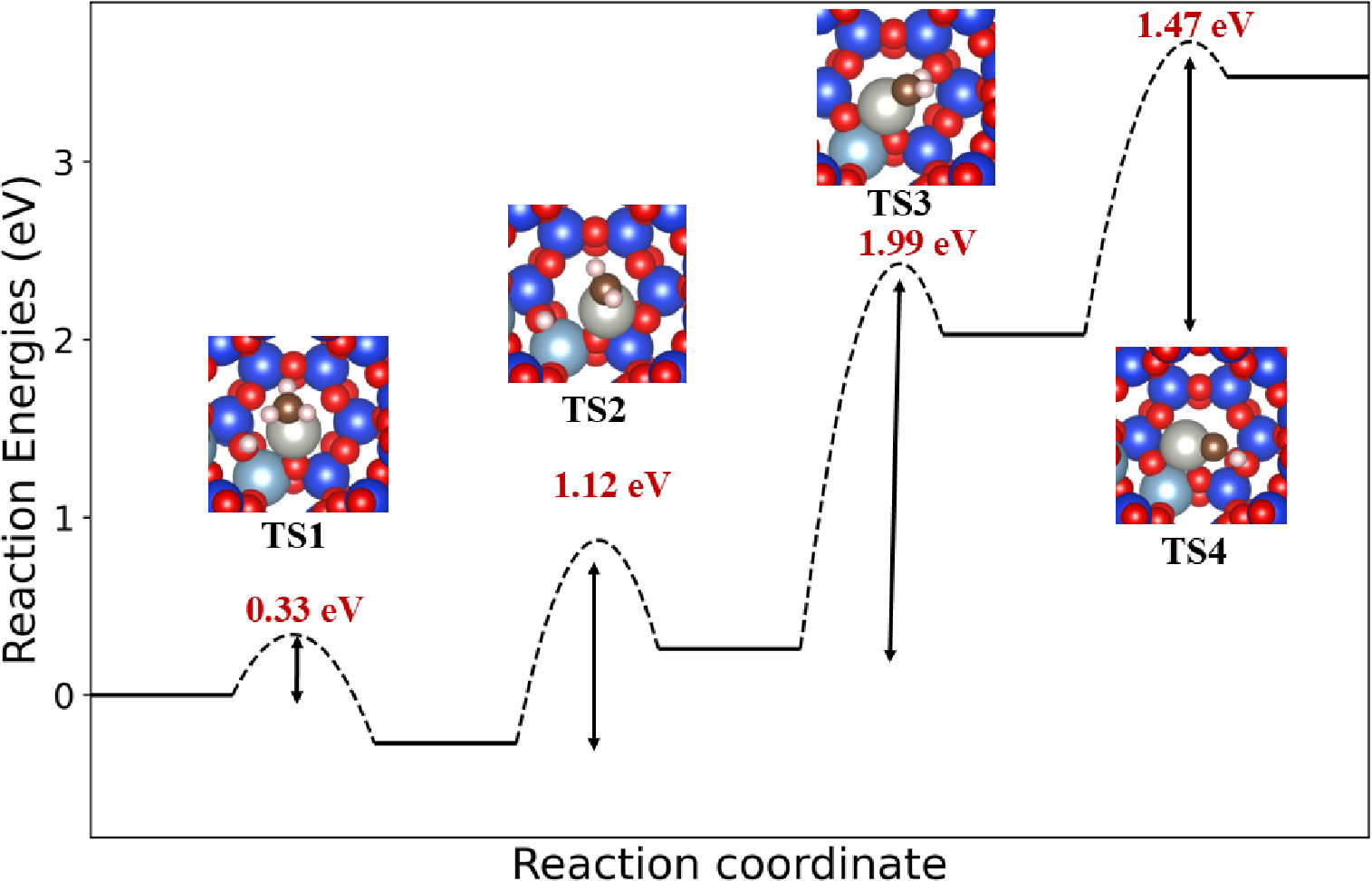}
\centering
\caption{(Color online) Activation barrier along with the transition states for the sequence under the direct dehydrogenation (dry)pathway $CH_4 \rightarrow CH_3 \rightarrow CH_2 \rightarrow CH \rightarrow C$  . The arrow indicate the magnitude of the activation barrier.}
\label{TS_dry_new}
\end{figure*}

\textit{Direct dehydrogenation of CH$_4$ on  Pd@SSZ-13: }
Direct dehydrogenation involves four important steps, i.e the conversion of CH$_4$ to  C intermediate species. The optimized structures are shown in Fig. \ref{Str_dry}(a).  Fig. \ref{TS_dry_new} provide the details of reaction energy profile of these four reaction steps along with the transition states and the corresponding equations are given below. 

$$ CH_4^* \rightarrow CH_3^* + H^* $$
$$ CH_3^* \rightarrow CH_2^* + H^* $$
$$ CH_2^* \rightarrow CH^* + H^* $$
$$ CH^* \rightarrow C^* + H^* $$

In the initial step, the physisorbed CH$_4$ molecule loses one of its hydrogen atom, resulting in the formation of CH$_3^*$. This reaction step involves a process change from physisorption to chemisorption, accompanied by C-H bond breaking and the C-Pd bond making. CH$_3^*$ prefers to move to the top-site of the Pd atom and the distance between C and Pd atom decreases from  2.37 \AA  to  2.01 \AA, which is consistent with the result of Xinyuan Bu et al \cite{BU2021111891}. This reaction is exothermic, with an enthalpy change of  -0.27 eV and an activation barrier of  0.33 eV.  Subsequent reaction steps involves the sequential removal of hydrogen atoms, resulting in the formation of CH$_3^*$, CH$_2^*$, CH$^*$, and C$^*$. The second step is endothermic, with a reaction energy of 0.54 eV and an activation barrier of 1.12 eV.  CH$_2^*$  tends to move away from the top-site of the Pd top-site and settle in the bridge site of Pd and framework O site, with the C-Pd bond length further decreasing from  2.01  to 1.97 \AA.  Next, CH$^*$ prefers to align itself towards the Pd, similarly to the  CH$_2^*$ intermediate as shown in Fig. \ref{Str_dry}a, and the C-Pd bond length decreases from 1.97 \AA to 1.87 \AA. The activation of the C-H bond in the CH$_2^*$ $\rightarrow$ CH step requires an activation barrier of 1.99, which represents a 0.40 eV increase compared to the previous step. In addition, C$^*$ prefers to be in the bridge site between Pd and framework O site as shown in Fig.  \ref{Str_dry}a, with a bond length decreasing from 1.87 \AA to 1.74 \AA. The final step in the direct dehydrogenation reaction requires an activation barrier of 1.47 eV. Overall, the dehydrogenation of CH$_4^*$ to C$^*$ is an endothermic process. 

\textit{O$_2$ dissociation:}
Following dehydrogenation, the molecular oxygen, O$_2$ first undergoes dissociation into O* species as shown in Fig \ref{Str_dry}c. The most stable site for adsorption of these dissociated species is on the top of Pd with an adsorption energy of -2.65 eV. Upon adsorption, the O-O bond elongates from its gas-phase bond length of 1.24 \AA to 2.17 \AA. indicating weakening of the bond pair prior dissociation. The dissociation of O$_2$ on Pd@SSZ-13 proceeds with an activation barrier of 2.5 eV.  The SSZ-13 environment plays a key role in stabilizing these atomic oxygen species, which serve as an active oxidants for subsequent reactions.

\begin{figure*}[t]
\centering
\includegraphics[width=14.0cm,keepaspectratio=true]{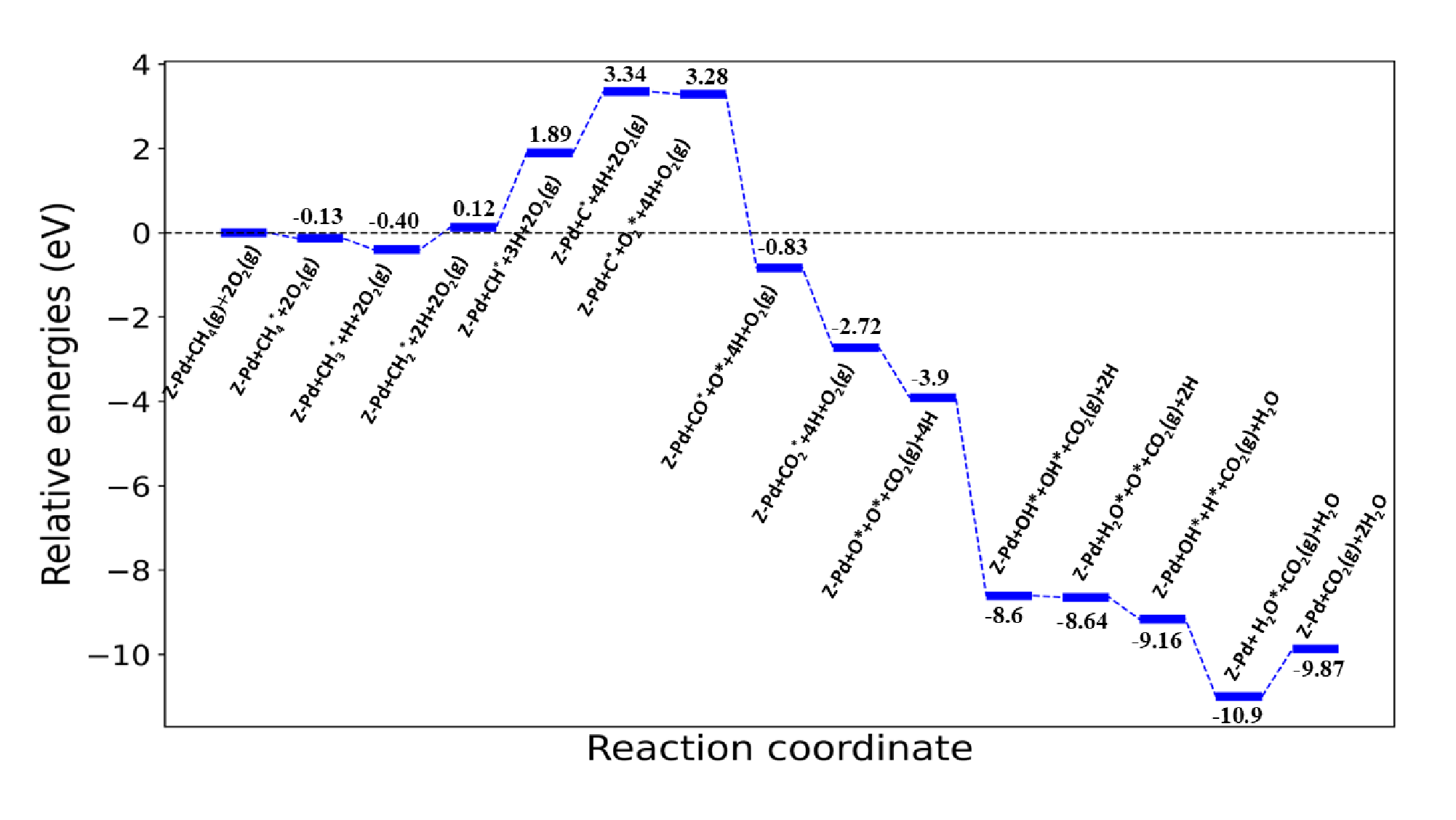}
\centering
\caption{(Color online) Complete reaction pathway for methane oxidation under dry condition over Pd@SSZ-13. In this context, (g) denotes species in the gas phase, while * indicates adsorbed species on the catalyst}
\label{RP_dry}
\end{figure*}

\textit{CO$_2$ and Water formation:}
The oxidation of surface-bound species on single-atom Pd@SSZ-13 proceeds via distinct pathways for carbon and hydrogen removal. Notably, direct O$_2$ dissociation on an isolated carbonaceous Pd site is not feasible due to the absence of Pd–Pd interactions that typically facilitate O–O bond cleavage \cite{Yu}. Instead, O$_2$ preferentially adsorbs on a Pd–C coadsorbed configuration, where the activated oxygen species readily react with the surface-bound C$^*$ to form CO$^*$(see Fig. \ref{Str_dry}b). This step is exothermic, with the confined microporous environment of SSZ-13 promoting enhanced orbital overlap between Pd, CO, and O$^*$ species, thereby facilitating CO$^*$ stabilization. Subsequent oxidation of CO$^*$ by an additional O$^*$ species yields CO$_2^*$, which then desorbs in a thermodynamically favorable process.
In parallel, water formation occurs through oxidation of surface H* species. Two competing pathways were considered: (i) OH$^*$ + H$^*$ $\rightarrow$ H$_2$O and (ii) OH$^*$ + OH$^*$ $\rightarrow$ H$_2$O + O$^*$. Among these, (i) is energetically more favorable pathway, requiring a lower barrier of 0.18 eV compared to the alternative route. Importantly, in the absence of coadsorbed carbon species, water formation proceeds exclusively through O$_2$ dissociation, highlighting a mechanistic distinction from the carbon oxidation sequence. The details of adsorption energies. reaction energies of important elementary reactions are given in Table \ref{tab:adsorption}, \ref{tab:Enthalpy} and a complete free energy profile summarizing both CO$_2$ and H$_2$O formation pathways is presented in Fig. \ref{RP_dry}.

\begin{table}[h]
\centering
\caption{Adsorption Energy ($E_{\text{ads}}$) of species involved in direct dehydrogenation (dry condition) and species involved in CO$_2$ and H$_2$O formation on the Pd@SSZ-13 catalyst}
\label{tab:adsorption}
\begin{tabular}{lcc lcc}
\hline
\textbf{species} & $E_{\text{ads}}$ (eV) & $d$ (\AA) & \textbf{species} & $E_{\text{ads}}$ (eV) & $d$ (\AA) \\
\hline
CH$_4^*$ & -0.13 & 2.37 & CO$^*$ + O$^*$        & -1.47 & 1.88 (CO$^*$), 1.835 (O$^*$) \\
CH$_3^*$ & -1.86 & 2.01 & CO$_2^*$       & -2.12 & 2.174  \\
CH$_2^*$ & -3.88 & 1.97 & OH$^*$         & -2.65 & 1.872 \\
CH$^*$   & -4.58 & 1.87 & H$_2$O$^*$     & -1.12 & 2.09 \\
C$^*$    & -5.21 & 1.74  & H$_2$O + O$^*$ & -1.28 & 2.122 (H$_2$O),1.78 (O$^*$) \\
\hline
\end{tabular}

\begin{flushleft}
\footnotesize{$^{a}$ $d$ represents the distance between the Pd atom and species in the optimized geometries as shown in Figures 1 and 2.}
\end{flushleft}
\end{table}

\begin{table}[h!]
\centering

\caption{Reaction Energies ($\Delta H$) for important elementary Reactions Involved in the dry condition (i.e Direct dehydrogenation and Oxidation Process)\textsuperscript{a}}
\label{tab:Enthalpy}
\begin{tabular}{l c l c}
\hline
\textbf{dehydrogenation} & $\Delta H$ & \textbf{oxidation} & $\Delta H$ \\
\hline
$* + CH_{4}(g) \rightarrow CH_{3}^* + H$ & $-0.27$ & $* + O_{2}(g) \rightarrow O^* + O^*$ & 1.93 \\
$CH_{3}^* \rightarrow CH_{2}^* + H$ & $0.53$ & $CO^* + O^* \rightarrow CO_{2}$ & -1.88 \\
$CH_{2}^* \rightarrow CH^* + H$ & $1.77$ & $OH^* + H \rightarrow H_2O^*$ & -1.83 \\
$CH^* \rightarrow C^* + H$ & $1.45$ & $OH^* + OH^* \rightarrow H_{2}O+O^*$ & -0.03\\

\hline
\end{tabular}

\textsuperscript{a}In this context, (g) and * represent species in the gas phase and adsorbed state, respectively.
\end{table}

\subsubsection{Wet/O2-assisted condition}

\textit{O$_2$-rich condition:}

\begin{figure*}[t]
\centering
\includegraphics[width=8.0cm,keepaspectratio=true]{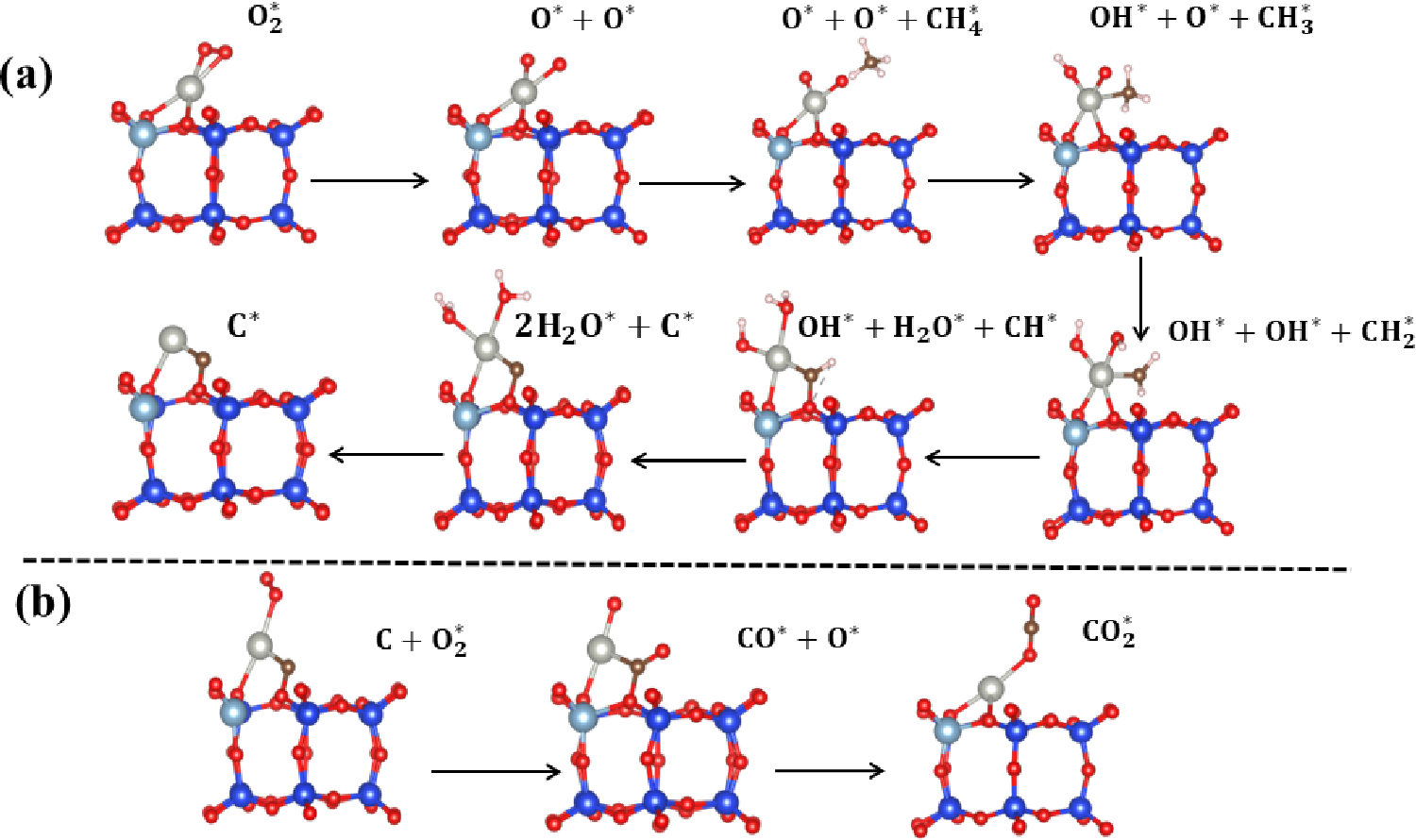}
\includegraphics[width=7.0cm,keepaspectratio=true]{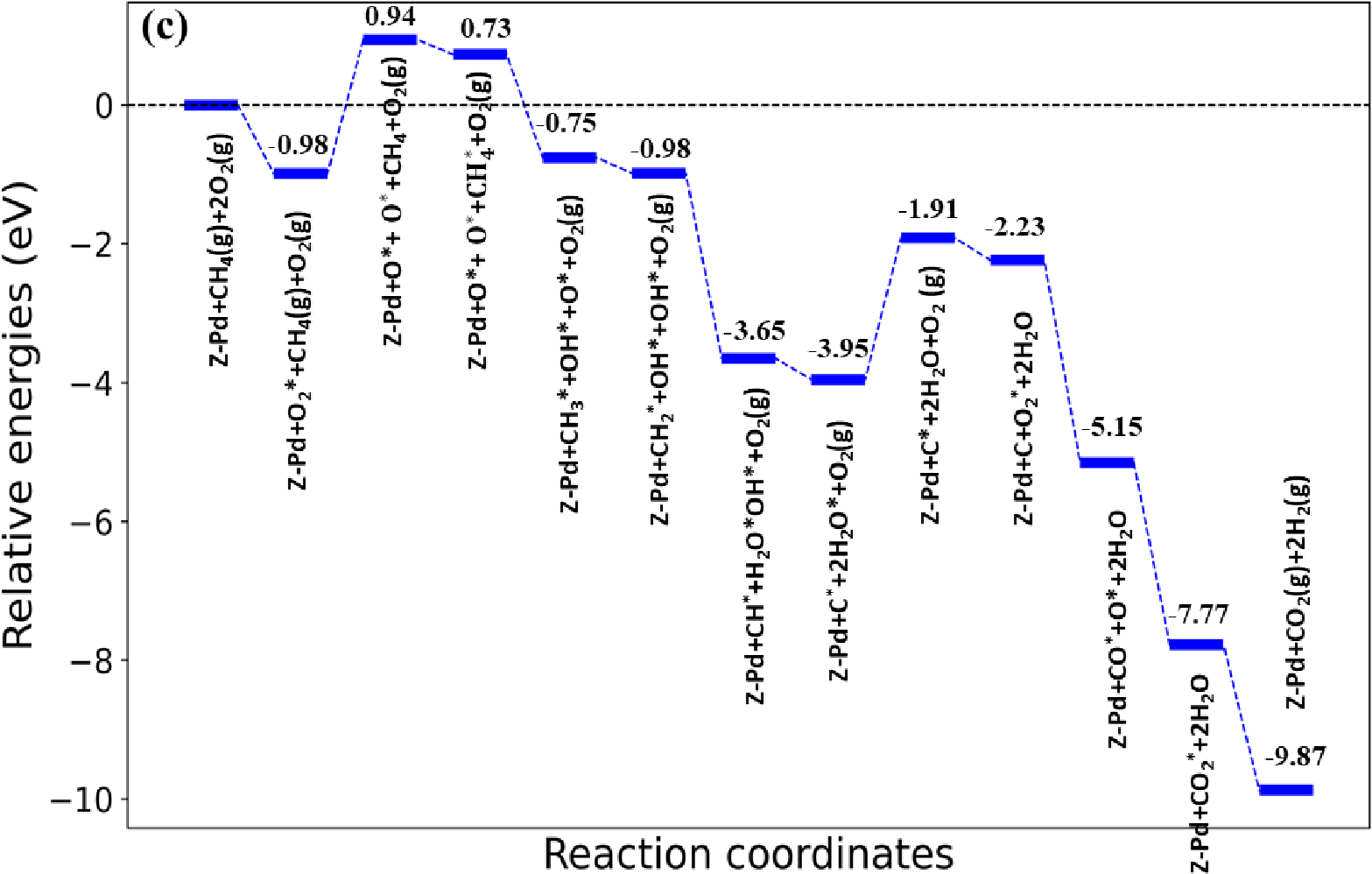}
\centering
\caption{(Color online) Complete reaction pathway for methane oxidation under O$_2$-rich condition over Pd@SSZ-13. (a-b) shows the representative optimized structures of key intermediates, while (c) represents the corresponding reaction energy profile. Here (g) denotes the gas-phase and * indicates adsorbed species on the catalyst}
\label{RP_O2rich}
\end{figure*}

To explore the thermodynamical feasibility of the methane oxidation over Pd@SSZ-13, a reaction energy profile was constructed  involving successive C–H bond scission of CH$_4$ leading to surface-bound carbon (C$^*$), followed by oxidation to CO$_2$. The optimized structures and the reaction energy profile for the complete methane oxidation on Pd@SSZ-13 has been shown in Fig. \ref{RP_O2rich}. This pathway initiates with the adsorption of O$_2$ on the Pd site with an adsorption energy of -0.98 eV. The O$_2$ molecule is  activated upon adsorption with the O-O bond length elongated from 1.24 to 1.29  \cite{Xue2021},   forming reactive lattice oxygen (O$^*$) capable of hydrogen abstraction. The two bond lengths of Pd-O become 2.001 and 2.01 \AA respectively. 
Upon adsorption of oxygen over Pd@SSZ-13, physisorption of the CH$_4$ molecule on a single atom Pd active site takes place and the methane molecule is pushed farther from the active site with the Pd-C(H4) distance, 3.97 A. This is subsequently followed by key elementary step , i.e C-H bond activation of CH$_4$ by lattice oxygen, resulting in the formation of a methyl (CH$_3^*$) intermediate and a hydroxyl (OH$^*$) group on Pd site (Z–Pd–O$^*$ + CH$_4$ $\rightarrow$ Z–Pd–OH$^*$ + CH$_3^*$).   The bond lengths of Pd-C(H3) and Pd-O(H) becomes 2.075 \AA and 1.919 \AA respectively. 
At this step, the O$_2$ dissociation occur with an O-O distance of 2.828 A, resulting in the formation of a methyl group and a hydroxyl group (OH*) which are bonded to the Pd site.  Further reaction steps involve the C-H bond breaking with the formation of  CH$_2^*$, CH$^*$, and C$^*$ intermediates as shown in Fig \ref{RP_O2rich}a. 
With the completion of methane activation, we proceed further with the reaction steps toward methane oxidation. Upon the formation of C* on the Pd site, as mentioned above, another O$_2$ molecule is introduced to the Pd@SSZ-13. This step leads to the adsorption of O$_2$ molecule over the Pd active site with the O-O bond length of 1.260, Pd-O bond lengths of 2.106 A, 2.767 A and Pd-C bond length of 1.861 A. The O-O bond length is found to be elongated when compared to that of free O$_2$ molecule, which is an indication of activation of O$_2$ over Pd. Now, the O atom which is away from the Pd atom gets tilted towards C atom for the formation of CO$^*$ with  a bond length of 1.188 A. This reaction step further undergoes with the formation of CO$_2$ by the bond formation of CO$^*$ with O$^*$ atoms over Pd site. Therefore, the complete CH$_4$ oxidation pathway over the Pd@SSZ-13 occurs via rich oxygen condition as follows


\begin{align*}
Z{-}Pd + O_2^* + CH_4(g) &\;\;\rightarrow\;\; Z{-}Pd + O_2^* + CH_4 \\
&\;\;\rightarrow\;\; Z{-}Pd + CH_3^* + OH^* + O^* \\
&\;\;\rightarrow\;\; Z{-}Pd + CH_2^* + OH^* + OH^* \\
&\;\;\rightarrow\;\; Z{-}Pd + CH^* + H_2O^* + OH^* \\
&\;\;\rightarrow\;\; Z{-}Pd + C^* + 2H_2O^* \\
&\;\;\rightarrow\;\; Z{-}Pd + C^* + 2H_2O \\
&\;\;\rightarrow\;\; Z{-}Pd + C^* + O_2(g) \\
&\;\;\rightarrow\;\; Z{-}Pd + CO^* + O^* \\
&\;\;\rightarrow\;\; Z{-}Pd + CO_2^*
\end{align*}

This pathway demonstrates the ability of Pd@SSZ-13 to facilitate complete oxidation of CH$_4$ through a dehydrogenation sequence that utilizes reactive oxygen species for both hydrogen abstraction and final oxidation of surface carbon. While thermodynamically favorable ($\Delta E$ = –9.87 eV), the stepwise formation of CH$_2^*$, CH$^*$, and C$^*$ introduces potential kinetic barriers and risks of carbon accumulation. Particularly, the H$_2O$ removal step represents a major kinetic and thermodynamic bottleneck in the complete oxidation of methane over Pd@SSZ-13. This step is markedly endothermic and leaves behind a surface-bound C* species, which serves as a potential coke precursor. The persistence of such carbonaceous intermediates can deactivate isolated Pd sites and obstruct zeolite micropores, ultimately reducing catalyst lifetime. These insights are consistent with experimental studies on Pd-zeolite systems, where deactivation is frequently attributed to carbon deposition and moisture-induced instability \cite{LIU2024128, GOPAL2002231, D3SC05851D, Zhou2024}. The strong correlation between theoretical and experimental findings underscores that optimizing the energetics of H$_2$O desorption is central to achieving stable and highly active Pd@SSZ-13 catalysts for methane combustion.

\textit{O$_2$-lean condition:}
\begin{figure*}[t]
\centering
\includegraphics[width=7.0cm,keepaspectratio=true]{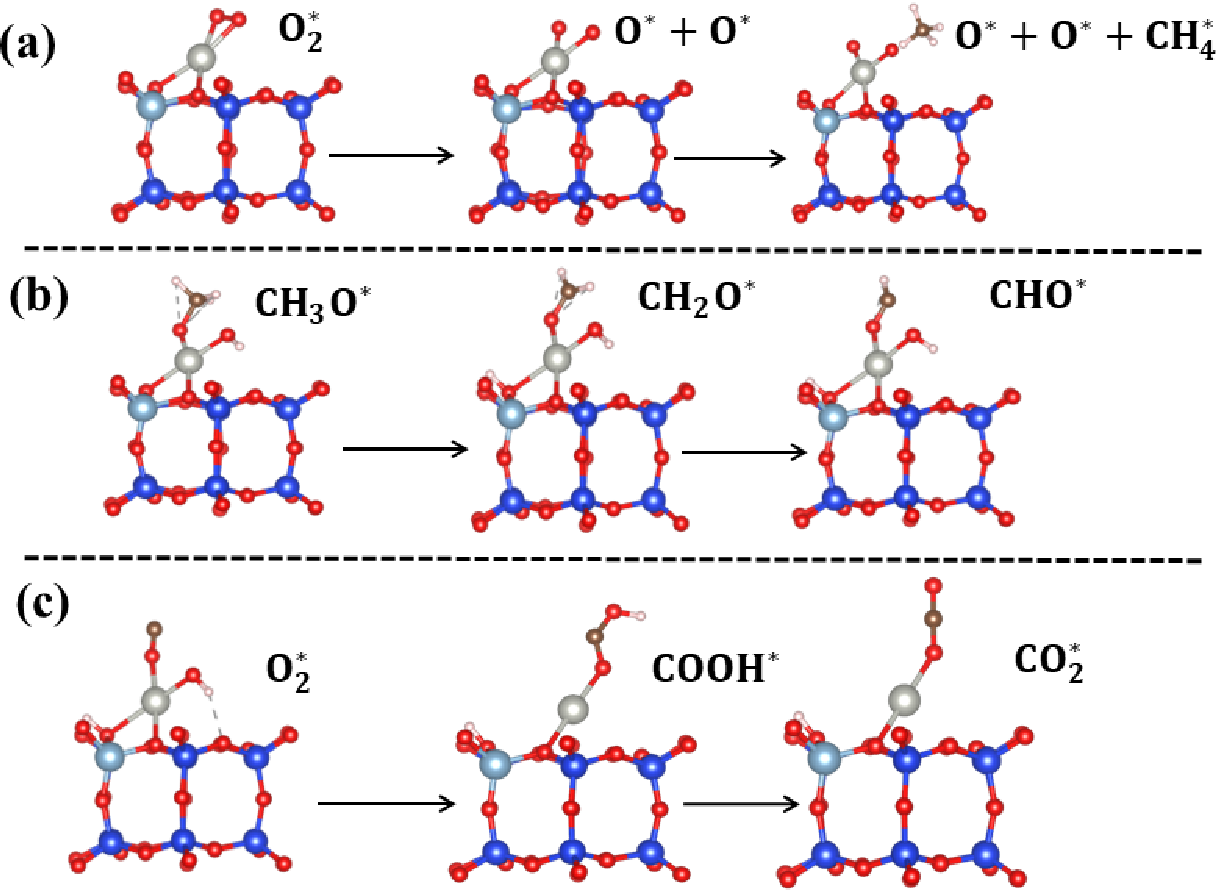}
\includegraphics[width=8.0cm,keepaspectratio=true]{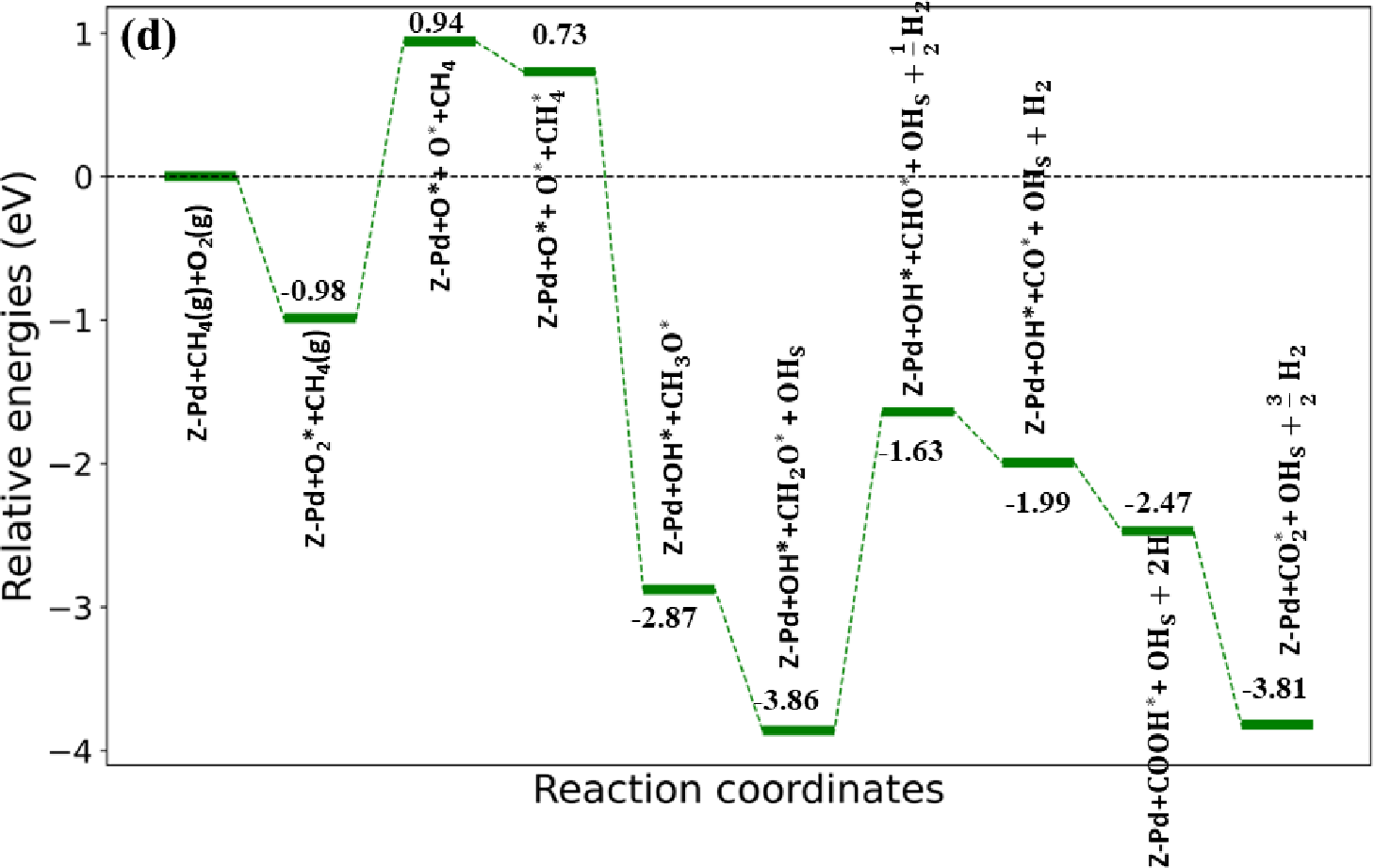}
\centering
\caption{(Color online) Complete reaction pathway for methane oxidation under O$_2$-lean condition over Pd@SSZ-13. (a-c) shows the representative optimized structures of key intermediates, while (c) represents the corresponding reaction energy profile. Here (g) denotes the gas-phase and * indicates adsorbed species on the catalyst}
\label{RP_O2lean}
\end{figure*}
To address the limitations observed in the previous pathway, particularly the uphill energy barrier associated with water desorption and the formation of surface-bound C* species, we explored an alternative reaction pathway for complete methane oxidation over Pd@SSZ-13, as shown in Fig \ref{RP_O2lean}. This pathway avoids the formation of C* species (coke precursor), which is identified as a potential bottleneck in the earlier mechanism. This new pathway initiates with the adsorption of O$_2$ as seen in O$_2$ rich condition.
In this new pathway, the  initial steps involving the O$_2$ adsorption and C-H bond activation mirror the previously outlined mechanism, highlighting the key role of lattice oxygen species (O$^*$).  Further route leads to the formation of surface-bound CH$_3$O$^*$ and CH$_2$O$^*$ intermediates instead of atomic C$^*$ species. The CH$_2$O$^*$ is subsequently transformed to CHO$^*$ with the removal of H, which is energetically unfavorable ($\Delta E = 2.22099$ eV).  Following this step, the CO$^*$ is formed, which then converts to COOH*, ultimately leading to the formation of CO$_2^*$. The final state, comprising CO$_2$, surface hydroxyl groups, and released hydrogen, is calculated to lie 3.81 eV lower than the energy of the isolated gas-phase reactants, confirming that the overall reaction is highly exergonic.
This O$_2$-assisted pathway highlights the Mars-van Krevelen-type mechanism, in which lattice oxygen from PdO-like species is consumed during CH$_4$ activation and is replenished by gas-phase O$_2$. The presence of the zeolite framework (SSZ-13) stabilizes the Pd species and likely contributes to the confinement effects that favor selective oxidation intermediates. The results emphasize the bifunctional catalytic nature of Pd@SSZ-13, where Pd sites mediate redox chemistry while the zeolite supports molecular adsorption and intermediate stabilization.

\subsection{Insights into the assessment of activation energy barriers: NEB vs semi-emphirical scaling relations}
Two distinct mechanistic routes for complete methane oxidation over Pd@SSZ-1were examined: dry pathway and oxygen-assisted pathway [O2-rich  \& O2-lean] to identify kinetically relevant steps. Although the CI-NEB method is widely regarded as the most accurate approach for explicit transition-state searches, we preferred the usage of semi-empirical scaling relations to reduce computational cost. 
The first method is based on the Brønsted--Evans--Polanyi (BEP) relationship \cite{BLIGAARD2004206}:
\begin{equation}
E_a = \gamma \, \Delta E + \xi
\end{equation}
where $\gamma$ is the BEP slope characteristic of a given reaction family, $\Delta E$ is the reaction energy, and $\xi$ is the intercept. The intercept was calibrated using a CI-NEB calculation of CH$_4$ activation over Pd@SSZ-13 in each case. Slopes were taken from literature values as follows $\gamma(C-H)$ =0.93, $\gamma(O-O)$=0.86, $\gamma(O-H)$=0.75, $\gamma(C-O)$=0.80 \cite{BEP_values, Wang2011, Vojvodic}
To cross-validate these findings, we applied the UBI-QEP method in the Shustorovich formulation,\cite{SHUSTOROVICH19981} as benchmarked by Bittencourt \textit{et al.} \cite{Bittencourt}. The activation energy is written as:  
\begin{equation}
E_a = \varphi \, \Delta E + \xi
\end{equation}
where $\varphi$ is the interpolation parameter reflecting transition-state character. Bittencourt \textit{et al.}  showed that tuning $\varphi$ to $\approx 0.85$ greatly improved agreement with NEB barriers across diverse surface reactions\cite{Bittencourt}. Following a similar strategy, we optimized these parameters for our system which is found to be $\varphi$=0.81, using CH$_4$ NEB barrier.

\subsubsection{Dry Pathway: Assessment of Activation Barriers}
In the dry condition, the computed reaction energy profile (Fig. \ref{RP_dry}(d)) shows a stepwise C–H bond cleavage sequence (CH$_4$ $\rightarrow$ CH$_3$ $\rightarrow$ CH$_2$ $\rightarrow$ CH $\rightarrow$ C), followed by oxidation of the carbonaceous intermediate. Among these, the dehydrogenation steps CH$_2$ $\rightarrow$ CH and CH $\rightarrow$ C exhibit the largest positive reaction energies ($\Delta E = 1.77$ and 1.45 eV, respectively), indicating their potential role as rate-determining steps.  We have calculated the barrier using CI-NEB for these steps and the transition states are given in Figure. \ref{TS_dry_new}. The O$_2$ dissociation step is also highly relevant with a computed NEB barrier of 2.50 eV, comparable to values reported for metal-zeolite and oxide-supported catalysts \cite{Shalini2023}.
For the direct dehydrogenation pathway, activation barriers were estimated using semi-empirical relations, the Brønsted–Evans–Polanyi (BEP) approach. This method was referenced to the  CH$_4$ $\rightarrow$ CH$_3$ step (NEB $E_a$= 0.33 eV, $\Delta E$=-0.27 eV). Initially, we adopted a literature-derived slope of $\Gamma$=0.93 (for C-H bond activation). However, this resulted in a poor correlation (low coefficient of determination, R$^2$). To improve the accuracy, the BEP slope has been subsequently refined and tuned for this system. With an adjusted slope of $\Gamma$=0.77, a significantly better fit was obtained. The detailed analysis and results are provided in SI (see Fig )  

Table~\ref{tab:activation_barriers_dry} summarizes the computed barriers. Both BEP and UBI-QEP predictions reproduce the NEB anchors by construction, and show consistent values (within $\sim$0.05 eV) across the C–H sequence and they are in reasonable agreement with the CI-NEB results. 

\begin{table}[h!]
\caption{Comparision of activation energy estimates (in eV) for key steps in the dry methane oxidation pathway over Pd@SSZ-13 using CI-NEB, BEP and UBI-QEP scaling anchored to in-system CI-NEB calculations.}
\label{tab:activation_barriers_dry}
\begin{tabular}{ccccc}
\hline
\textbf{Step} & \textbf{$\Delta E$ (eV)} &\textbf{$E_a$ (CI-NEB)} &\textbf{$E_a$ (BEP)} & \textbf{$E_a$ (UBI-QEP)} \\
\hline
CH$_4$ $\rightarrow$ CH$_3$ & -0.27 & 0.330 &0.330 & 0.330  \\
CH$_3$ $\rightarrow$ CH$_2$ & 0.53 & 1.12  &0.978 & 0.981 \\
CH$_2$ $\rightarrow$ CH & 1.77 & 1.99 & 1.902 & 1.990 \\
CH $\rightarrow$ C & 1.45 & 1.47  &1.654 & 1.730 \\
\hline
\end{tabular}
\end{table}

\subsubsection{Oxygen-assited Pathway: Assessment of Activation Barriers}
In the Oxygen-rich (blue) pathway, the largest positive reaction free energy corresponds to water removal from the Z-Pd+C$^{*}$+2H$_{2}$O$^{*}$ intermediate ($\Delta E = 2.04097$ eV). However, the rate-determining step is formally defined by the highest activation barrier as shown in Table~\ref{tab:activation_barriers}. In the oxygen-lean (green) pathway, the most endergonic step involves the hydrogen abstraction from CH$_2$O$^{*}$ to form CHO$^{*}$ and H$^{*}$ ($\Delta E = 2.22099$ eV), likewise corresponds to the highest activation barrier reported in  Table~\ref{tab:activation_barriers}.
Hence, these steps were considered as rate-determining steps (RDS) for subsequent kinetic barrier estimation, consistent with prior computational studies on hydrocarbon oxidation over metal-zeolite catalysts\cite{Shalini2023}. 
Table~\ref{tab:activation_barriers} summarizes the computed activation barriers for the two rate-determining steps using both BEP and UBI-QEP approaches. The two methods are in close agreement, differing by less than 0.02 eV, confirming the reliability of the semi-empirical scaling relations.  

\begin{table}[h!]
\caption{Activation energy estimates (in eV) for the identified rate-determining steps in the oxygen-rich (blue) and oxygen-lean (green) pathways.}
\label{tab:activation_barriers}
\begin{tabular}{lccc}
\hline
\textbf{Pathway (RDS)} & \textbf{$\Delta E$ (eV)} & \textbf{$E_a$ (BEP)} & \textbf{$E_a$ (UBI-QEP)} \\
\hline
Oxygen-rich (blue): H$_2$O removal & 2.041 & 2.064 & 2.070 \\
Oxygen-lean (green): CH$_2$O$^{*}$ $\rightarrow$ CHO$^{*}$ + H$^{*}$ & 2.221 & 2.251 & 2.216 \\
\hline
\end{tabular}
\end{table}

 It is observed that, in the dry pathway, the reaction is hindered by substantial barriers associated with C–H bond activation and O$_2$ dissociation. Both the BEP and UBI-QEP semi-empirical relationships, validated against the dry case, consistently capture these trends. Using these tested approaches, we estimated the barriers for the oxygen-assisted route. Notably, both methods predict that the oxygen-rich pathway exhibits a lower kinetic barrier (by $\sim$0.19 eV), indicating that this route is more favorable compared to the Oxygen-lean pathway, provided that coke formation is mitigated.
The barrier landscape in Tables~\ref{tab:activation_barriers_dry}--\ref{tab:activation_barriers} now feeds directly into a unified rate/TOF analysis (Sec.~\ref{subsec:kinetics_tof}), followed by a full microkinetic treatment (Sec.~\ref{sec:microkinetics}) that resolves coverages and rate control under operating $(T,p)$.
\subsection{Kinetics and Turnover Frequencies: from Barriers to Rates}
\label{subsec:kinetics_tof}

Having established activation barriers for the dry and O$_2$-assisted pathways (Tables~\ref{tab:activation_barriers_dry} and \ref{tab:activation_barriers}), we map these values onto rate constants and apparent turnover frequencies (TOFs). Throughout we adopt the transition-state-theory prefactor
\[
A = \frac{k_{\mathrm B} T}{h},
\]
with barrier heights interpreted as Gibbs barriers $\Delta G^{\ddagger}$ at the simulation temperature (enthalpic barriers plus vibrational and configurational corrections when available). The elementary rate for step $i$ is therefore
\begin{equation}
 k_i(T) = A\,\exp\!\left[-\frac{\Delta G_i^{\ddagger}}{k_{\mathrm B}T}\right].
\label{eq:TST}
\end{equation}
To connect with coverage and gas-phase conditions, we evaluate the site-normalized production rate of CO$_2$ as
\begin{equation}
 \mathrm{TOF} = r_{\mathrm{CO_2}} = k_{\mathrm{RDS}}(T)\,\prod_j a_j^{\nu_j}\,\theta_{\mathrm{RDS}},
\label{eq:TOF_general}
\end{equation}
where $a_j=p_j/p^{\circ}$ are activities of gas-phase reactants, $\nu_j$ are the stoichiometric orders entering the rate-determining step (RDS), and $\theta_{\mathrm{RDS}}$ is the coverage of the surface intermediate that undergoes the RDS. Coverages are obtained from Langmuir adsorption using the same adsorption free energies that feed the microkinetic model in Sec.~\ref{sec:microkinetics}. Equation~\eqref{eq:TOF_general} subsumes the simplified $k(T)\,\theta$ form used previously; pressure effects now enter explicitly through the activities $a_j$.

\begin{figure*}[t]
\centering
\includegraphics[width=8.0cm,keepaspectratio=true]{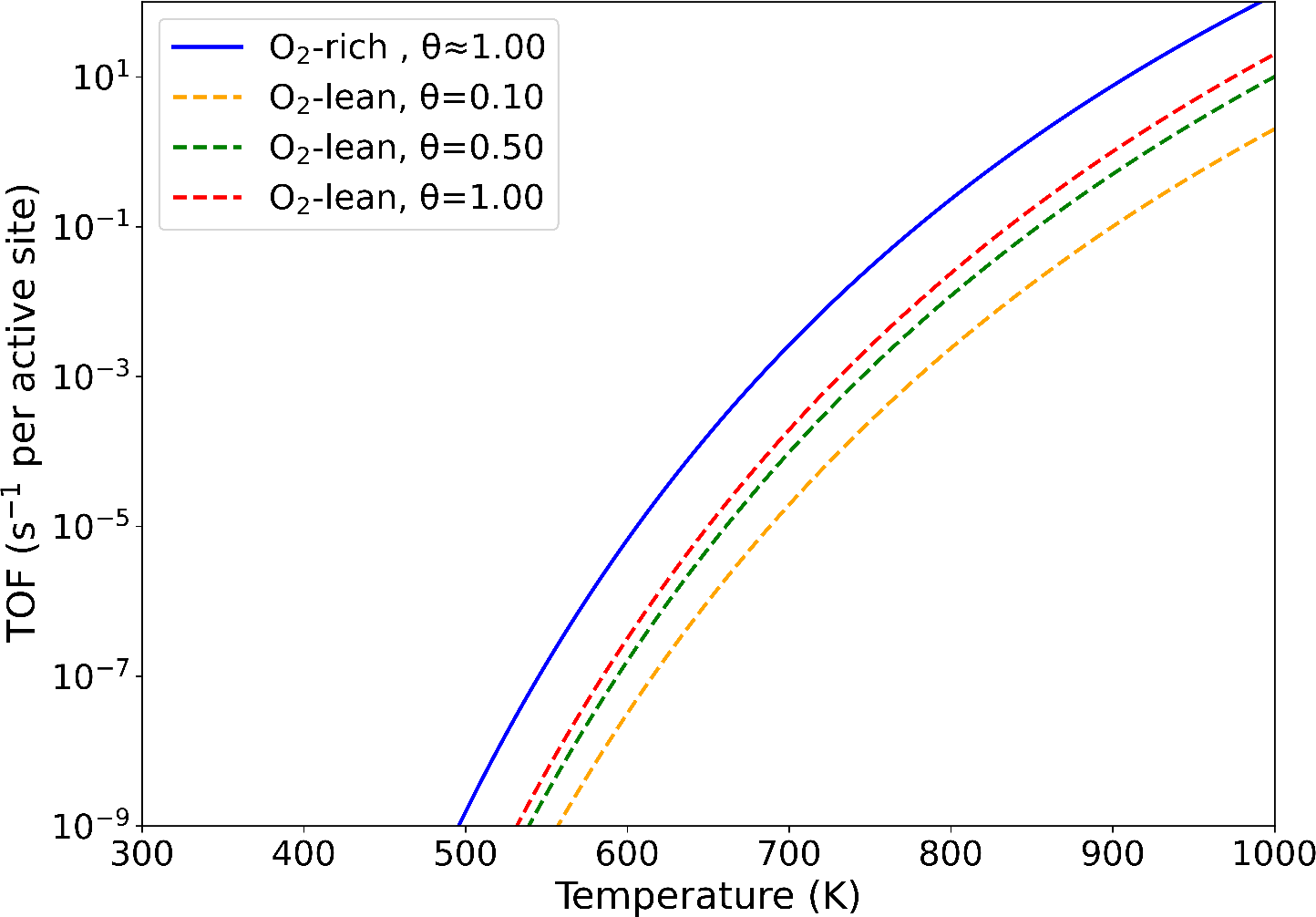}
\centering
\caption{(Color online) Turnover frequency (TOF) profiles for methane oxidation under O$_2$-rich ($\theta$ = 1.0) and O$_2$-lean conditions ($\theta$ = 0.1, 0.5, 1), highlighting the influence of oxygen coverage on catalytic performance.}
\label{TOF}
\end{figure*}

\subsubsection{O\texorpdfstring{$_2$}{2}-rich pathway} With the RDS identified as H$_2$O removal (Table~\ref{tab:activation_barriers}), we evaluate $k_{\mathrm{RDS}}(T)$ from Eq.~\eqref{eq:TST} using $\Delta G^{\ddagger}=2.06$~eV. The surface coverage $\theta_{\mathrm{RDS}}\equiv\theta_{\mathrm{H_2O^*}}$ follows from the adsorption free energy of H$_2$O on Pd@SSZ-13; at 700~K and $p(\mathrm{H_2O})=0.01$~bar we obtain near-saturation ($\theta \approx 1$). This yields a site-normalized TOF of $\mathcal{O}(10^{-3})$~s$^{-1}$, consistent with the pathway-level profiles in Fig.~\ref{TOF}. Despite the favorable rate constant at high $T$, this route passes through a deep C$^*$ well, so kinetic advantage must be balanced against deactivation risk (Sec.~\ref{RP_O2rich}).

\subsubsection{O\texorpdfstring{$_2$}{2}-lean pathway}
Here the RDS is CH$_2$O$^*\rightarrow$CHO$^*+\,\mathrm{H}^*$ with $\Delta G^{\ddagger}\approx 2.22$~eV. Strong binding keeps $\theta_{\mathrm{RDS}}\simeq\theta_{\mathrm{CH_2O^*}}\approx 1$ across wide $p$--$T$ windows, so the intrinsic barrier dominates. The resulting TOF at 700~K is $\mathcal{O}(10^{-4})$~s$^{-1}$ (Fig.~\ref{TOF}). Although slower, this route avoids the C$^*$ accumulation that challenges the oxygen-rich pathway.

Figure~\ref{TOFPT} decomposes the global TOF trends into elementary-step propensities (CH$_4$ C--H activation, O$_2$ dissociation, H$_2$O formation) across $(T,p)$ using Eq.~\eqref{eq:TST} and the activities in Eq.~\eqref{eq:TOF_general}. The sharp rise of O$_2$-dissociation TOF only at the high-$T$/high-$p$ corner reflects its large barrier; the broad basin of H$_2$O-formation TOF stems from lower barriers combined with high $\theta_{\mathrm{OH^*}/\mathrm{H^*}}$ once activation has occurred. Together with Fig.~\ref{TOF}, these maps clarify why the blue pathway becomes rate-competitive only after oxygen activation is facile and why water removal remains the kinetic bottleneck thereafter.

All symbols are now harmonized across the kinetics and microkinetic analyses: $k_{\mathrm B}$, $A=k_{\mathrm B}T/h$, activities $a_j=p_j/p^\circ$, and site coverages taken from the shared adsorption dataset. We next connect these ingredients to the full microkinetic treatment in Sec.~\ref{sec:microkinetics}.

\subsubsection{Pressure--temperature TOF maps for key elementary steps}
\label{subsec:tof_maps}
\begin{figure*}[t]
\centering
{\includegraphics[trim=0.00cm 0.00cm 0.00cm 0.00cm,clip=true,width=1.0\textwidth]{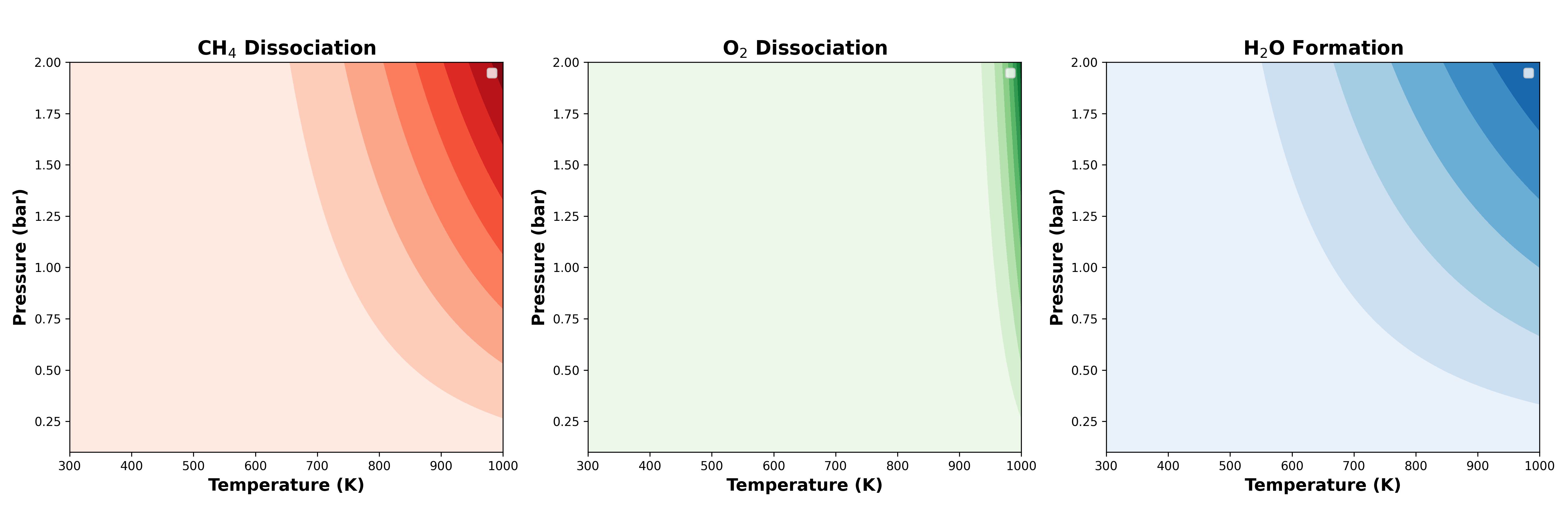}}
\centering
\caption{The turnover frequency (in s$^{-1}$) of CH$_4$ and O$_2$ dissociation and H$_2$O formation concerning temperature (in K) and pressure (in bar).}
\label{TOFPT}
\end{figure*}

Using Eq.~\eqref{eq:TST} with barrier inputs from Tables~\ref{tab:activation_barriers_dry} and \ref{tab:activation_barriers}, and activities $a_j=p_j/p^\circ$, we computed step-resolved TOF fields over $T\in[300,1000]$~K and $p\in[0.1,2]$~bar. The resulting heatmaps in Fig.~\ref{TOFPT} show: (i) CH$_4$ C--H activation strengthens with $T$ and moderate $p$; (ii) O$_2$ dissociation is negligible except at the high-$T$/high-$p$ corner, consistent with its large barrier; and (iii) H$_2$O formation is broadly facile once activated intermediates accumulate. These fields are not stand-alone rates but the ingredients entering the pathway-level TOFs in Fig.~\ref{TOF}. Quantitative TOF values quoted above use the same prefactor and coverages as the microkinetic model in the following section below.

\subsection{Microkinetic Modelling}
\label{sec:microkinetics}

The microkinetic modeling was carried out using the MKMCXX software \cite{Filot}. All adsorbed species were assumed to occupy a single adsorption site. For surface reactions, the forward and reverse rate constants were determined using the Eyring equation,
\begin{equation}
 k = \frac{k_{\mathrm B}T}{h}\,\exp\!\left(-\frac{\Delta G^\ddagger}{k_{\mathrm B}T}\right),
\end{equation}
where $k$ is the reaction rate constant, $\Delta G^\ddagger$ is the Gibbs free energy barrier, $k_{\mathrm B}$ is the Boltzmann constant, $h$ is Planck's constant, and $T$ is the absolute temperature. The following expressions were used to calculate adsorption and desorption rates with gas-phase driving forces expressed as activities $a_j=p_j/p^\circ$,
\begin{equation}
   K_{\mathrm{ads}}  = \frac{a_j A}{\sqrt{2\pi m k_{\mathrm B}T}}\,S,
\end{equation}

\begin{equation}
   K_{\mathrm{des}}  = \frac{k_{\mathrm B}T^3}{h^3}   \frac{A(2\pi m k_{\mathrm B})}{\sigma \Theta_{\mathrm{rot}}}\,\exp\!\left(-\frac{E_{\mathrm{des}}}{k_{\mathrm B}T}\right),
\end{equation}

where $A$ is the adsorption site area, $m$ is the relative molecular mass, $S$ is the sticking coefficient, $\sigma$ is the symmetry number, $\Theta_{\mathrm{rot}}$ is the rotational characteristic temperature, and $E_{\mathrm{des}}$ is the desorption energy. Activities reduce to partial-pressure ratios for ideal gases, so $a_j=p_j/p^\circ$. The value of symmetry number ($\sigma$) for CH$_4$, O$_2$, CO$_2$, H$_2$O is 12, 2, 2, 2 \cite{donald}, and the corresponding rotational temperatures ($\Theta_{\mathrm{rot}}$) are 7.54 K, 1.47 K, 0.56 K, and 20.9 K, respectively \cite{atkins, adriana, donald}. For simplicity, the sticking coefficients were assumed to be unity in all microkinetic simulations \cite{Jiao, Jianlin}. The apparent activation energy was then evaluated using the following relation

\begin{equation}
  r =  k_{\mathrm{ads}}\,a_A -k_{\mathrm{des}}\,[A_{ads}],
\end{equation}

\begin{equation}
  E_{\mathrm{app}} = R T^2 \frac{d \ln r}{dT} = -R\frac{d \ln r}{d \frac{1}{T}},  
\end{equation}
where $r$ is the total reaction rate and $T$ is the temperature.

\begin{figure*}[t]
\centering
{\includegraphics[trim=0.00cm 0.00cm 0.00cm 0.00cm,clip=true,width=0.8\textwidth]{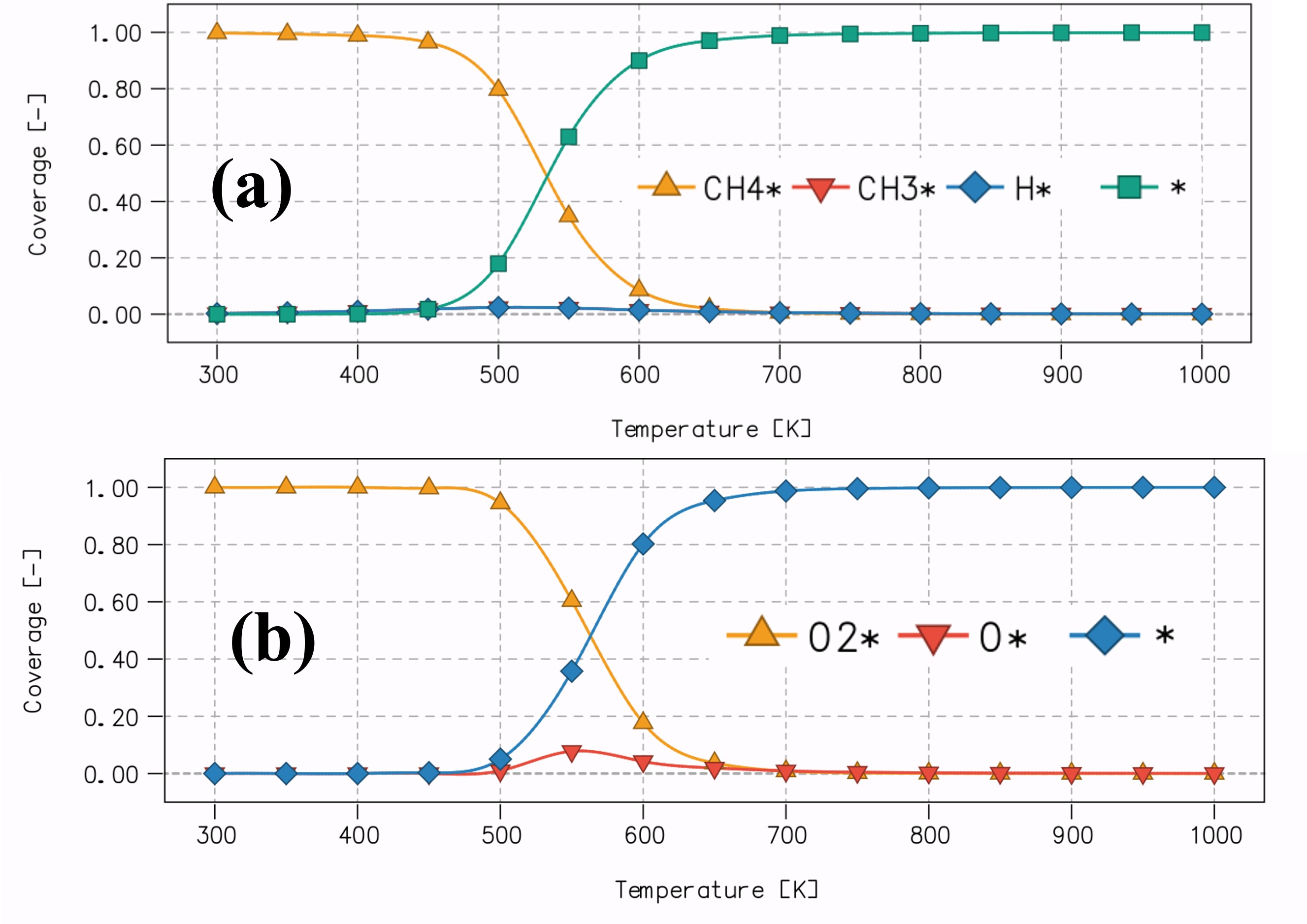}}
\centering
\caption{(a) The coverage of carbon species in reaction mechanism of CH$_4$ dissociation and  (b) oxygen species during O$_2$ dissociation with respect to temperature 300-1000 K at pressure 1 bar.}
\label{DissoMC}
\end{figure*}

Figure~\ref{DissoMC}(a) shows that at low temperatures (300-450 K), the surface is almost completely saturated with CH$_4^*$ species, while free sites are nearly absent and only negligible amounts of CH$_3^*$ and H$^*$ are observed, indicating that methane remains largely molecularly adsorbed. As the temperature increases to the intermediate range (450-600 K), CH$_4^*$ coverage decreases sharply due to dissociation, while the number of free sites rises rapidly, approaching full availability around 600 K. In this regime, small amount of CH$_3^*$ and H$^*$ species appear, indicating the beginning of methane activation.  At higher temperatures ($>$ 600 K), the surface becomes dominated by free sites, with methane-derived intermediates nearly vanishing. For oxygen species, the surface is initially saturated with O$_2^*$, leaving no free sites and only minor amounts of O$^*$, reflecting strong oxygen adsorption as represented in Figure~\ref{DissoMC}(b) . As the temperature rises to 500–600 K, O$_2^*$ coverage decreases sharply, free sites increase and dominate by $\approx$ 650 K, and a small amount of O$^*$ appears, indicating O$_2$ dissociation. At higher temperatures ($>$ 650 K), the surface becomes completely clean, with free sites approaching unity and no oxygen adsorbates remaining. This trend highlights that oxygen strongly blocks the surface at low temperature but dissociates at elevated temperature, thereby regenerating active sites for further reactions.

\begin{figure*}[t]
\centering
{\includegraphics[trim=0.00cm 0.00cm 0.00cm 0.00cm,clip=true,width=0.8\textwidth]{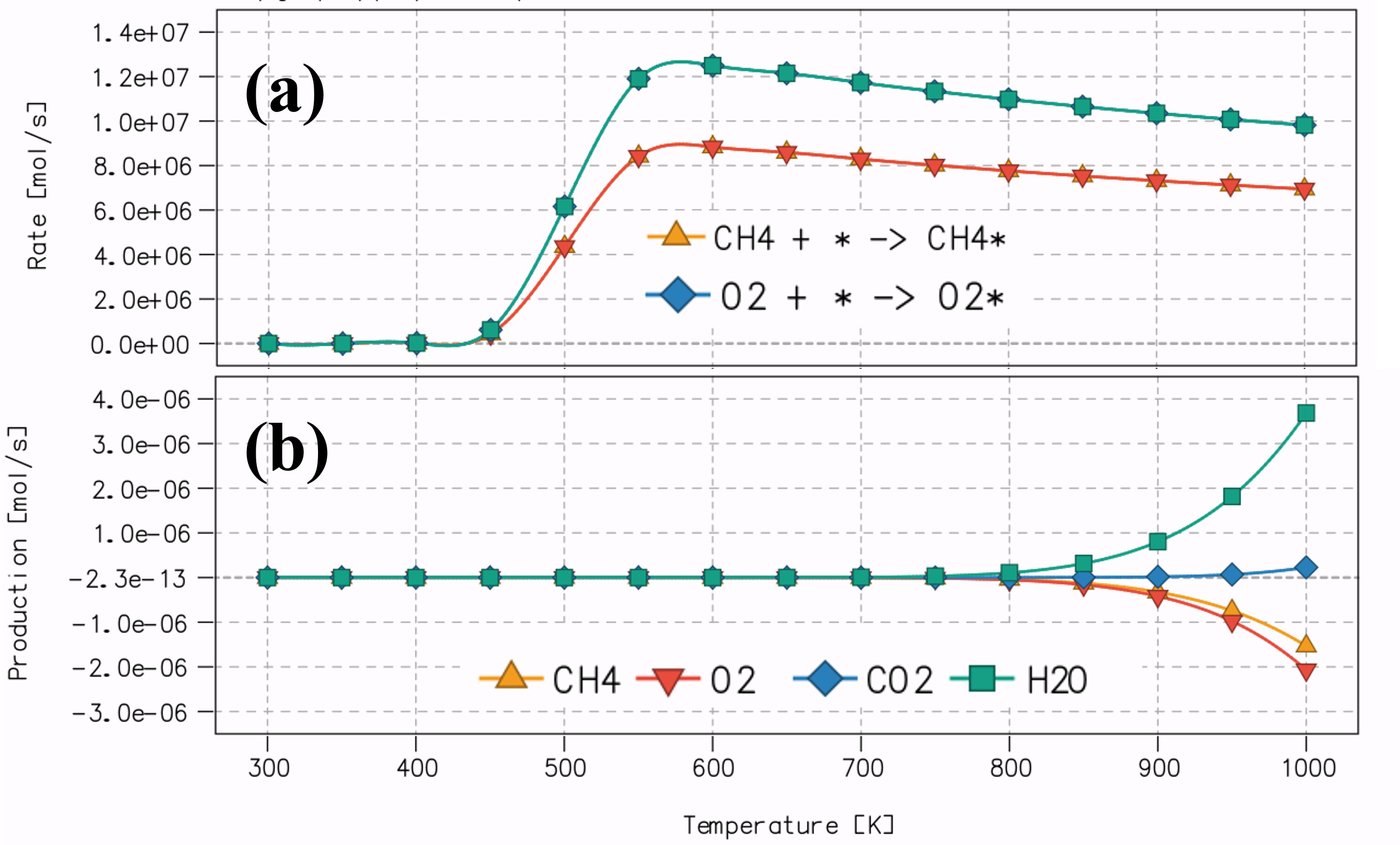}}
\centering
\caption{(a) The rate of adsorption of CH$_4$ and O$_2$ and (b) the production rate of CO$_2$ and H$_2$ from the decomposition of CH$_4$ and O$_2$ with respect to temperature 300-1000 K at pressure 1 bar.}
\label{RateMC}
\end{figure*}

The adsorption rates of CH$_4$ and O$_2$ remain negligible at low temperatures (300–450 K) but as the temperature increases to the intermediate range (450–650 K), the adsorption rates rise sharply. The rates peaks around 550–600 K, consistent with the surface transition observed in Fig.~\ref{RateMC}(a). Above 650 K, adsorption decreases because molecules do not stick well, while desorption takes over, leaving the surface mostly clean. Thus, the main catalytic activity occurs between 500–650 K, where adsorption is most active. Fig.~\ref{RateMC}(b) shows the production rates of CO$_2$ and H$_2$O resulting from the conversion of CH$_4$ and O$_2$. In the range of 300–600 K, product formation remains minimal, but as the temperature increases to 600–800 K, both H$_2$O and CO$_2$ start to form, with water emerging as the dominant product and the consumption of CH$_4$ and O$_2$ also starting.
Above 800 K, H$_2$O production rises sharply and stays dominant, CO$_2$ increases more gradually, and CH$_4$ and O$_2$ consumption becomes faster with temperature. Overall, the reaction shifts to full oxidation at higher temperatures, producing H$_2$O and CO$_2$ as the main products.

\begin{figure*}[t]
\centering
{\includegraphics[trim=0.00cm 0.00cm 0.00cm 0.00cm,clip=true,width=0.8\textwidth]{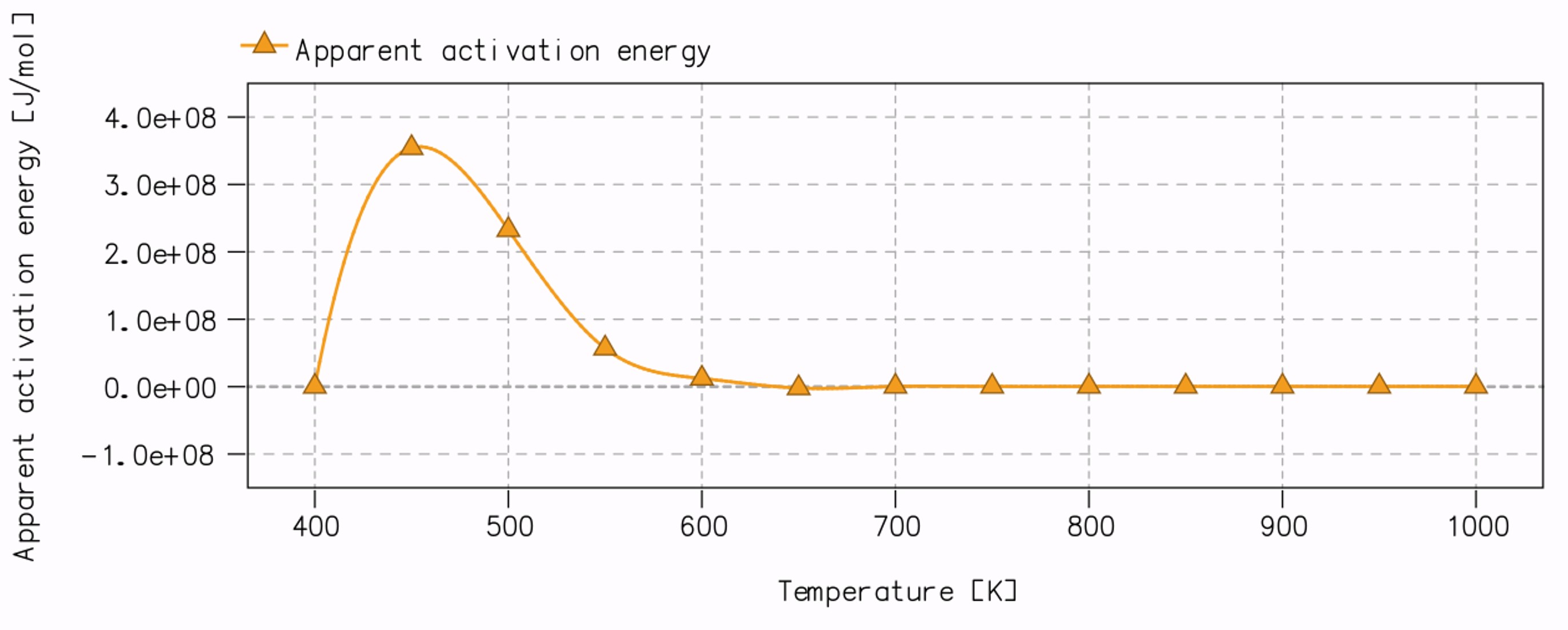}}
\centering
\caption{The apparent activation barrier as a function of the temperature (in K) for the
methane oxidation into CO$_2$ and H$_2$O 400-1000 K at pressure 1 bar.}
\label{AB}
\end{figure*}

The apparent activation energy (E$_{app}$) is near zero around 400 K due to surface blockage by CH$_4^*$ and O$_2^*$, with the rate governed by adsorption equilibrium, as deicted in Fig.~\ref{AB}. Between 450–550 K, E$_{app}$, app rises sharply (3–4 $\times$ 10$^8$ J/mol) as desorption and bond-breaking steps dominate, marking the activation-controlled regime. Above 600 K, E$_{app}$ drops and stabilizes at a low value, indicating a transport or adsorption limited regime with a nearly temperature-independent reaction rate.
\newpage

\newpage
\section{Conclusion}\label{secIV}
In summary, the first principles simulations were carried out to understand the mechanistic pathways of complete methane oxidation over Pd single-atom catalyst incorporated into SSZ-13 zeolite. All possible Pd incorporated configurations were analyzed systematically for finding the  most stable active site configuration. Comparative analysis of the direct (dry) and O$_2$-assisted oxidative dehydrogenation pathways revealed that the dry pathway is both kinetically and thermodynamically unfavorable, whereas the presence of activated O$_2$ substantially lowers C–H bond activation barriers and promotes an overall exothermic reaction profile. However, the calculated reaction energy profile highlights that the H$_2$O removal step is a critical bottleneck, given its endothermic characteristic. This reaction step results in the retention of surface-adsorbed carbon species, which could potentially act as precursors to coke formation. Such findings align with experimental data on the deactivation of Pd/zeolite catalysts, where catalyst longevity is compromised by carbon accumulation and sensitivity to moisture. 
Overall, this work highlights that the inherent C–H activation capability of palladium and the energy required for H$_2$O removal, are critical determinants of catalyst stability and activity. These insights establish a mechanistic foundation for the deliberate development of palladium/zeolite catalysts that are resistant to moisture and coke formation.

\begin{acknowledgement}
This work was supported financially by the Fundamental Research Program (PNK9750) of the Korea Institute of Materials Science (KIMS), South Korea. The authors also would like to acknowledge the IKST cluster facility for computing time.   
\end{acknowledgement}

\bibliography{manuscript}

\end{document}